\documentclass[prl,twocolumn,showpacs,amsmath,amssymb]{revtex4}
\usepackage{graphicx} 
\usepackage{dcolumn} 
\usepackage{bm} 
\def \bea{\begin{eqnarray}}
\def \eea{\end{eqnarray}}
\begin{document}


\title{Stripes, Zigzags, and Slow Dynamics in Buckled Hard Spheres}

\author{Yair Shokef}
\thanks{Present Address: Physics of Complex Systems, Weizmann Institute of Science, Rehovot 76100, Israel.}
\author{Tom C. Lubensky}
\affiliation{Department of Physics and Astronomy, University of
Pennsylvania, Philadelphia, PA 19104, USA}
\date{\today}

\begin{abstract}

We study the analogy between buckled colloidal monolayers and the triangular-lattice Ising antiferromagnet. We calculate free volume-induced Ising interactions, show how lattice deformations favor zigzag stripes that partially remove the Ising model ground-state degeneracy, and identify the Martensitic mechanism prohibiting perfect stripes. Slowly inflating the spheres yields jamming as well as logarithmically slow relaxation reminiscent of the glassy dynamics observed experimentally. 

\end{abstract}



\maketitle


Geometric frustration, manifested for example in the
anti-ferromagnetic (AF) Ising model on a triangular lattice
\cite{Wannier}, occurs whenever local interaction energies cannot be
simultaneously minimized. It gives rise to highly degenerate ground
states, unusual phases of matter \cite{Moessner06}, and possibly
slow or glassy dynamics \cite{Tarjus}, whose properties even after
decades of research are not fully understood. Here we present a
theoretical study of a colloidal system, one of a class
of artificial frustrated systems in which the state of each constituent can be
directly visualized \cite{art_frus}, that provides new insight into
the microstructure of frustrated systems and its connection with
their dynamics.

We study hard spheres confined between
parallel plates.  For plate separation slightly greater than the
sphere diameter and at sufficiently high sphere density,
the spheres buckle upward or downward \cite{Pieranski83,buckling,hs}
from their lower-density positions on a hexagonal lattice.  This
buckling gives rise to a choice of two states for each sphere,
analogous to the two states of Ising-model spins
\cite{Pieranski83,Ogawa83}.  The tendency of spheres to maximize
free volume introduces an effective repulsive interaction that
favors configurations with neighboring spheres in opposite states
just as the AF Ising interaction favors opposite states of
neighboring spins. As in the triangular-lattice AF Ising model,
frustration arises because it is impossible to arrange the three
particles on any triangular plaquette such that all
pairs of neighboring particles are in opposite states.  In the AF
Ising model on a rigid lattice, there is an extensive number of
ground-state configurations (implying an extensive ground-state
entropy) in which neighboring spins on two of the three bonds on
each plaquette are in opposite states.  Given the analogy just drawn
between our colloidal system and the AF Ising model, it is
reasonable to conjecture that the colloidal system might exhibit
ground-state degeneracies and dynamics similar to those of the
rigid-lattice AF Ising model. Recent experiments 
in diameter-tunable-microgel systems
revealed sub-extensive ground-state entropy and glassy dynamics \cite{Han08}. 
Our theoretical study will address the differences between
the colloidal system and the rigid lattice Ising model and make
some conjectures about a likely closer analogy between the colloidal
system and the AF Ising model on a compressible lattice \cite{elas}.

We show that the short-ranged AF behavior in this
hard-sphere system may be explained by a simple geometrical model
relating it to the nearest-neighbor Ising model. However, the
out-of-plane buckling induces local in-plane lattice distortions
that, as in the elastic Ising model, partially remove the Ising
ground-state degeneracy and select configurations with zigzagging
stripes of up and down spheres. This `ground-state' lacks the local
zero-energy modes found in the Ising model on a rigid triangular
lattice, and as a result, the colloidal system exhibits dynamics
that are qualitatively slower than those of the
Ising model. Moreover, stripes require global deformations that are
incompatible with the system's boundary conditions; consequently
they break up into a Martensite \cite{Bhattacharya} and form
a new partially disordered and highly degenerate `ground-state'.


The free energy of our hard-sphere system is dictated by its phase-space
volume, which is a collective function of all the particles in the
system; hence this system is not exactly equivalent to an Ising
model with pairwise additive interactions. Nonetheless, we may
compare our system to the nearest-neighbor Ising model on the
triangular lattice and ask what is the strength of
AF interactions that best describes the hard-sphere
system.

A `microscopic' state is specified by the positions $\{ x_i, y_i, z_i
\}$ of all particles $1 \leq i \leq N$. We coarse-grain these states
into Ising-like configurations specified by $\{ s_i \}$ with $s_i =
{\rm sign}(z_i)$ ($x,y$ are the coordinates in the plane of the
confining walls, $z$ is perpendicular to the walls, and $z=0$ is at
the middle of the cell). The probability of a particular
configuration $\{s_i\}$ is equal to
the $3N$-dimensional integral $V\left(\{s_i\}\right)$ over all
states belonging to that configuration, divided by
the total phase-space volume $V_{\rm tot}$ of all configurations:
\bea
P_{\rm HS}\left(\{s_i\}\right) = V\left(\{s_i\}\right) / V_{\rm tot} .
\eea
We would like to equate this to the probability of finding the
corresponding configuration in the Ising model,
\bea
P_{\rm I}\left(\{s_i\}\right) = \exp\left(\beta J \sum{s_i s_j}
\right) / Z ,
\eea
where $\beta=1/k_BT$ is the inverse temperature, $J$ is the
interaction strength, the sum runs over all nearest neighbor pairs,
and $Z$ is the canonical partition function.

\begin{figure}[t]
\includegraphics[width=0.869\columnwidth]{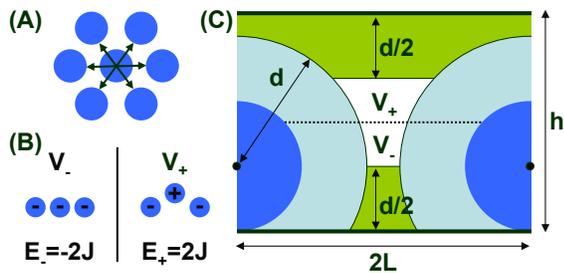}
\caption{(Color online) Free volume model: A) Top view.
Contributions to free volume originate from motion along axes to
each of the neighbors. B) Side view. Up particle surrounded by down
particles has more free volume (lower energy) than a down particle.
C) Surrounding neighbors touch wall and are separated by $2L$,
central particle is confined to the vertical plane. Large circles
are volumes excluded by neighbors, horizontal bands are volumes
excluded by the walls, remaining white region is the central
particle's free volume, divided at the middle of the cell height
(dotted line) into $V_{\rm +}$ and $V_{\rm -}$.}
\label{fig:free_vol_model}
\end{figure}

Unlike the commonly used cell-model, which approximates the
phase-space volume of a system as a product of single-particle free
volumes, our model equates $P_{\rm HS}$
and $P_{\rm I}$ by assuming that $V\left(\{s_i\}\right)$ is a
product of contributions from all nearest-neighbor `bonds':
$V\left(\{s_i\}\right) = \prod{v\left(s_i s_j ; h, d, L \right)}$ ,
where the pair contribution $v$ depends on $s_i s_j$ and on
the wall separation $h$, the sphere diameter $d$, and the in-plane
number density, which we characterize by the spacing $L$ of the
underlying triangular lattice. We evaluate $v(s_i s_j)$ in a
quasi-one-dimensional approximation by allowing particles $i$ and
$j$ to move only in the vertical plane passing through the axis
connecting their lattice positions (see
Fig.~\ref{fig:free_vol_model}A). We consider a particle, which we
call the central particle, and its two neighbors along one of the
principal lattice directions (see Fig.~\ref{fig:free_vol_model}B).
If the two neighbors are in opposite Ising states, the free volumes
resulting from the central one being up or down are equal by
symmetry. When the two neighbors are in the same state (down,
without loss of generality), the central particle has more free
volume ($V_{\rm +}$) when it is up than when it is down ($V_{\rm
-}$). We calculate $V_{\rm +}$ and $V_{\rm -}$ from the geometrical
setting of Fig.~\ref{fig:free_vol_model}C and equate the ratio of
the probabilities of finding the two configurations in
Fig.~\ref{fig:free_vol_model}B for hard spheres to that of the Ising
model: 
\bea 
\frac{V_{\rm +}}{V_{\rm -}} = \frac{\exp(-\beta E_{\rm
+})}{\exp(-\beta E_{\rm -})} = \exp(-4 \beta J).
\eea 
From this,we deduce that the hard-sphere system corresponds to an
Ising model with an effective AF interaction, 
$\beta J_{\rm eff}(d,h,L) = - \ln \left( V_{\rm +} / V_{\rm -} \right) / 4 < 0 $.

\begin{figure}[t]
\includegraphics[width=\columnwidth]{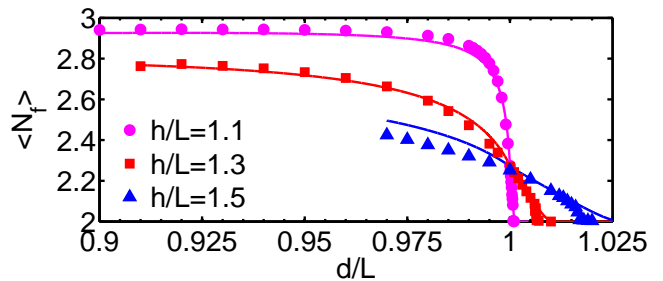}
\caption{(Color online) Average number of frustrated bonds per
particle vs sphere diameter and cell height. Free-volume model
(lines) agrees with Monte-Carlo simulations (symbols).}
\label{fig:map_to_ising}
\end{figure}

For given geometrical parameters $d$, $h$, $L$, we evaluate $V_{\rm
+}$ and $V_{\rm -}$ and determine the effective interaction strength
$\beta J_{\rm eff}$. We then use the exact solution of the Ising
model \cite{Wannier} to calculate the average number $\langle N_f
\rangle$ of frustrated neighbors per particle (we refer to
$s_is_j=-1$ as satisfied and to $s_is_j=1$ as frustrated), which
provides a measure of short-range AF order. $\langle N_f \rangle =2$
is the value in the ground state, and $\langle N_f \rangle =3$
corresponds to a random configuration. Figure~\ref{fig:map_to_ising}
shows the agreement between our simple model and three-dimensional
Monte-Carlo (MC) simulations. Simulations included $N=1600$ spheres
with steps consisting of small displacements of single spheres as
well as area-preserving box deformations, in which the angle or
aspect ratio of the simulated parallelogram was allowed to change
\cite{hs}. To probe cases with $d>L$, we start with striped
configurations that can accommodate the maximal sphere diameter by
lattice deformation (see below), wait $4$$\times$$10^5$ steps, and
then average $\langle N_f \rangle$ over additional $4$$\times$$10^5$
steps. We plot results only of cases for which $d$ was large enough
for the system to have long-range six-fold orientational order
$\Psi_6 \equiv \langle \exp(i6\theta_{jk}) \rangle > 0.5$
($\theta_{jk}$ is the angle the bond between $j$ and $k$ forms with
an arbitrary axis, and the average is over all
nearest-neighbors pairs \cite{2Dmelting}). Small spheres ($d<L$) in
a wide cell ($h/L=1.5$) are weakly confined, and the approximation
that the surrounding spheres in Fig.~\ref{fig:free_vol_model}C touch
the walls fails, giving rise to small differences between the model
and simulations.

\begin{figure}[t]
\includegraphics[width=0.941\columnwidth]{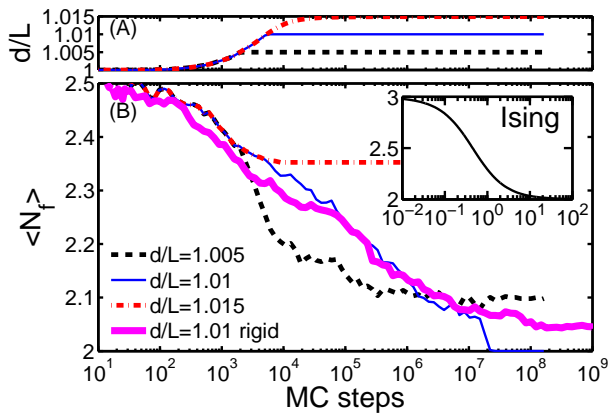}
\caption{(Color online) A) Sphere diameter, and, B) average number
of frustrated bonds per particles following swelling to different
sphere diameters. Normalized cell height $h/L=1.3$. Inset: same for
Ising model following quench to $T=0$.}
\label{fig:dynamics}
\end{figure}

For large spheres ($d>L$), simulations remain jammed in striped
configurations and do not increase the value of $\langle N_f
\rangle$ beyond the initial value of $2$, even though they are
expected to do so from the free volume considerations incorporated
in the model. To further explore this jamming, we conducted MC
simulations that started at a disordered configuration with $d_0=L$
and then ordered as the sphere diameter was gradually increased to
some larger value $d$. The spheres were initially on a triangular
lattice in the $xy$ plane with each sphere randomly touching either
the top or bottom wall. To speed the simulations, we considered
random jumps in the $z$ direction between touching either walls,
while keeping the $xy$ displacements continuous. During the swelling
process, once every MC step the diameter of all spheres was
increased to the maximal value allowable without overlaps.
Figure~\ref{fig:dynamics} shows results of simulations with wall
separation $h/L=1.3$. For $d/L = 1.005$, the free-volume model
predicts $\langle N_f \rangle = 2.12$, and the simulation indeed
slowly equilibrates to that value by $\sim$$10^5$ MC steps. For $d/L
= 1.01$, the system's relaxation to the value of $\langle N_f
\rangle = 2$ predicted by the model includes a logarithmically slow
decay to $\langle N_f \rangle \approx 2.1$ over a time scale of
$10^7$ MC steps, followed by a sharp jump to the equilibrium state.
For $d/L=1.015$, although the system is expected to be at a state
with $\langle N_f \rangle = 2$, it gets jammed during the swelling
process at a state with $\langle N_f \rangle = 2.35$ and does not
leave it over the time scales investigated here. Note that in
Fig.~\ref{fig:dynamics} we plot a single realization for each case,
however we observed similar behavior when repeating the simulations
with multiple realizations. Neither jamming nor logarithmically slow
relaxation occur in the Ising model even when quenched to zero
temperature (see inset).

\begin{figure}[t]
\includegraphics[width=0.969\columnwidth,trim=59 0 3 90,clip]{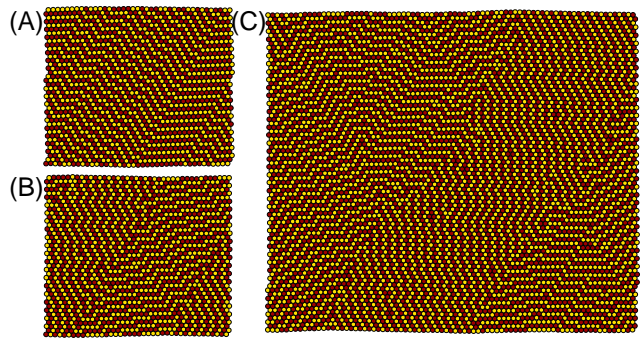}
\caption{(Color online) Final configurations following swelling to
$d/L=1.01$ at $h/L=1.3$. A) Deformable box. B-C) Rigid box. System
has $N=$ 1600(A-B), 6400(C) spheres. Spheres touching top/bottom
wall are dark/bright. Simulation boxes are deformed parallelograms
with periodic boundary conditions. For ease of presentation we copy
the simulated region and plot a rectangular region of the periodic
system.}
\label{fig:confs}
\end{figure}

\begin{figure}[t]
\includegraphics[width=\columnwidth]{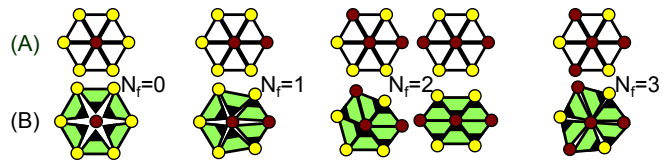}
\caption{(Color online) Tiling with: A) equilateral triangles for
the Ising ground-state, and, B) isosceles triangles for close-packed
buckled hard spheres. The large angle $\beta$ is blackened.}
\label{fig:triangles}
\end{figure}

Densely-packed spheres exhibit slower dynamics than low-temperature
Ising spins on a rigid lattice because the morphology of the
maximally-packed hard-sphere configurations differ from those of the
Ising ground-state. Unlike the highly disordered Ising ground-state
\cite{Wannier}, the hard-sphere `ground-state' consists of parallel
zigzag stripes (Fig.~\ref{fig:confs}A). In the Ising model, each
triangular plaqutte has one bond frustrated and two satisfied.
Although one third of the bonds in the system are frustrated, and
the average number of frustrated neighbors per particle is $\langle
N_f \rangle=2$, not all particles have exactly two frustrated
neighbors. By considering the six triangles surrounding a certain
spin in the lattice, Fig.~\ref{fig:triangles}A shows the five
possible ways (up to rotations and spin inversions) to align them
such that each triangle will have a single frustrated bond. The
central spin may have $N_f= 0$, $1$, $2$, or $3$ frustrated
neighbors. This leads not only to disorder but also to fast
relaxation dynamics since spins with $N_f=3$ are free to flip
without an energetic cost. For close-packed buckled spheres, each
triplet of spheres in contact defines an equilateral triangle with
sides $d$. As in the Ising model, one of the three spheres is up (or
down) and two down (or up), thus tilting this equilateral triangle
with respect to the horizontal plane. When projected onto the plane,
the tilted equilateral triangle is deformed to an isosceles triangle
with one long side $d$ along the frustrated bond and two shorter
sides $x=\sqrt{d^2-(h-d)^2}<d$ along the satisfied bonds. Each of
these isosceles triangles has two small angles
$\alpha=\cos^{-1}\left(\frac{d}{2x}\right)<\frac{\pi}{3}$ and a
large angle $\beta=\pi-2\alpha>\frac{\pi}{3}$. Now, close-packed
configurations for the buckled spheres are equivalent to tiling the
plane with these isosceles triangles. To completely cover the plane,
the angles of the six triangles meeting at each vertex must sum to
$2\pi$. Figure~\ref{fig:triangles}B demonstrates that for
$N_f=0,1,3$ the angles sum to $6\beta>2\pi$, $2\alpha+4\beta>2\pi$,
and $6\alpha<2\pi$, respectively, and thus that the triangles
cannot fit together: for the two
configurations with $N_f=2$ the angles sum to
$4\alpha+2\beta=2\pi$, enabling a perfect tiling corresponding to
the maximal-density close-packed state.

The slow dynamics observed for large sphere diameters result from
the lower degeneracy of these zigzagged stripe configurations
compared to the Ising ground-state. More importantly, the
close-packed states with $N_f \equiv 2$ do not have the free
particles with $N_f=3$ that are crucial for the low-temperature
dynamics in the Ising model \cite{kim}. Here, many spheres need to
cooperatively rearrange in order for the system to find
configurations that maximize the free volume. Spheres swollen to a
very large diameter ($d/L=1.015$ here) hardly move vertically to
change their Ising-configuration because the neighboring
spheres do not have enough room to rearrange in the horizontal
directions and to accommodate the lattice deformations required to
achieve optimal packing.


When the spheres swell slowly enough, they find a configuration that
maximizes free-volume by each sphere having exactly two frustrated
neighbors. Such configurations consist of parallel zigzagging
stripes (Fig.~\ref{fig:confs}A). Stripes run only along two of the
three principal lattice directions, hence the local distortions are
non-isotropic and require a macroscopic deformation of the system.
This is possible in the simulations described above in which the
shape of the simulation's bounding-box changes dynamically
\cite{hs}. However, experimentally the spheres form crystalline
domains separated by grain boundaries \cite{2Dmelting}, which may be
better described theoretically by rigid boundary conditions. Then,
the local tendency for zigzag stripes is incompatible with the rigid
boundary conditions. The tiling rules for the isosceles triangles
induce local deformations along two of the three principal lattice
directions, and for the system to be globally isotropic it must
break up into domains with stripes running along different
directions. We suspect that this is the
mechanism leading to the broken stripes seen experimentally
\cite{Han08} and we indeed observed such Martensitic states
\cite{Bhattacharya} when repeating the swelling simulations without
allowing the simulation box to deform \cite{footnote_weiss}. For instance, the case of
$h/L=1.3$, $d/L=1.01$ relaxes in the deformable box simulations to
the zigzagged striped state with $N_f \equiv 2$, whereas in a rigid
box, the average number of frustrated neighbors relaxes to $\langle
N_f \rangle = 2.05$ (Fig.~\ref{fig:dynamics}B), and the final
configuration (Fig.~\ref{fig:confs}B) consists of broken stripes.
We saw similar structures
(Fig.~\ref{fig:confs}C) and relaxation to the same value of $\langle
N_f \rangle$ for $N=100,400,1600,6400$. It would be interesting to test whether the size of these domains scales as the square root of the system size, as was found in other Martensites \cite{Bhattacharya}, and whether subsequently $\langle N_f \rangle$ slowly goes to 2 in the large-$N$ limit.


The maximal sphere diameter possible in a zigzag configuration is
equal to that of straight stripes, and the free-volume-cell
approximation does not distinguish between the two $N_f=2$
configurations corresponding to a straight segment of a stripe and
to a bend in the stripes. However simulations and experiments seem
to indicate a possible preference for straight stripes over zigzags.
It is unclear if the observed zigzag patterns represent equivalence
between straight and zigzagged stripes, or whether the system falls
into zigzagged configurations due to kinetic reasons. It would be
interesting to go beyond the mean-field description, as was done when
comparing the face-centred cubic and the
random hexagonal-close-packed structure of hard spheres in three dimensions
\cite{rhcp}.

The relief of frustration by lattice deformation resembles the
elastic Ising model \cite{elas}, which when analyzed exactly at the
microscopic level yields by our isosceles tiling scheme a
zigzag-stripe ground-state. It would be interesting to further
investigate the finite temperature behavior of that model, as well
as other models with zigzag-stripe ground-states \cite{Nussinov}.


\begin{acknowledgments}

We thank Yilong Han, Matt Lohr, Arjun Yodh, and Peter Yunker for involving us
in their experimental study of this topic, and Bulbul Chakraborty,
Randy Kamien, Andrea Liu, Carl Modes, Yehuda Snir, and Anton Souslov
for helpful discussions. This work is supported by NSF MRSEC grant
DMR-0520020.

\end{acknowledgments}


\end{document}